\documentstyle{europhys}
\def\And{{\rm and\ }}

\newif\ifboo \boofalse

\input epsf.sty
\begin{document}
\euro{}{}{}{}
\Date{}
\shorttitle{H. Falakshahi et al: Effect of a lattice}  

\title{Effect of a lattice upon an interacting system of electrons: \\\
Breakdown of scaling and decay of persistent currents} 

\author{H. Falakshahi $^{(a)}$, Z. \'A. N\'emeth$^{(a,b)}$ \And J.-L. 
Pichard$^{(a,c)}$} 

\institute{
(a) CEA/DSM, Service de Physique de l'Etat Condens\'e, \\
Centre d'Etudes de Saclay, F-91191 Gif-sur-Yvette Cedex, France \\  
(b) E\"otv\"os University, Departement of Physics of Complex Systems, \\
H-1117 Budapest, P\'azm\'any P\'eter s\'et\'any 1/A, Hungary \\
(c) Laboratoire de Physique Th\'eorique et Mod\'elisation, 
Universtit\'e de Cergy-Pontoise, 95031, Cergy-Pontoise Cedex, France 
}

\pacs{
\Pacs{71}{10-w} {Theories and models for many-electron systems}
\Pacs{71}{10.Fd} {Lattice fermion models} 
\Pacs{73}{20.Qt} {Electron solids}}

\maketitle

\begin{abstract}  For an interacting system of $N$ electrons, 
we study the conditions under which a lattice model of size $L$ 
with nearest neighbor hopping $t$ and $U/r$ Coulomb repulsion has 
the same ground state as a continuum model. For a fixed value 
of $N$, one gets identical results when the inter-electron spacing 
to the Bohr radius ratio $r_s < r_s^*$.  Above $r_s^*$, the persistent 
current created by an enclosed flux begins to decay and $r_s$ 
ceases to be the scaling parameter. Three criteria giving similar 
$r_s^*$ are proposed and checked using square lattices.
\end{abstract} 
 
 To numerically study an interacting system of $N$ electrons, it 
is convenient to use a lattice model of size $L$, which can 
be exactly solved when $N$ and $L$ are small. To compare exact 
lattice results with those obtained assuming a continuum space and 
hence unavoidable approximations (truncations, perturbative expansions, 
variational approaches, fixed node Monte Carlo 
calculations\cite{ceperley},...), one needs to know the role of the lattice 
upon the interacting system. Using the inter-particle spacing to the 
Bohr radius ratio $r_s$, we define in this letter the value $r_s^*$ 
which characterizes the onset of the lattice effects for the ground state 
(GS). Assuming that $N$ is given and varying the other parameters of the 
lattice model, one finds that below $r_s^*$ the lattice GS is identical 
to the continuum GS, $r_s$ is the scaling parameter, and 
the persistent current $I$ created by an enclosed flux keeps its value 
without interaction. Above $r_s^*$, the $N$ electrons are localized by 
$N$ lattice sites, the GSs become different, $I$ begins to decay, and 
a lattice physics where $r_s$ is no longer the scaling parameter takes 
place. 

 The Hamiltonian $H_c$ describing  $N$ electrons free to move 
on a continuum space of dimension $d$ contains one body kinetic 
terms, two body interaction terms plus a constant term 
due to the presence of the uniform positive background which can be 
added in order to have charge neutrality. Measuring the energies in 
rydbergs ($1 Ry = me^4/2\hbar^2$) and the lengths in units of the 
radius $a$ of a sphere (circle in $2d$) which encloses on the average 
one electron, $e$ and $m$ being the electronic charge and mass, 
$H_c$ reads
\begin{equation}
H_c=-\frac{1}{r_s^2} \sum_{i=1}^N \nabla_i^2 + \frac{2}{r_s}  
\sum_{1\leq i < j\leq N} \frac{1}{|{\bf r}_i-{\bf r}_j|} + const, 
\label{H-continuous}
\end{equation}
which only depends on the scaling ratio $r_s=a/a_B$ when $N \rightarrow 
\infty$. The Bohr radius $a_B=\hbar^2/me^2$ characterizes the scale for 
the quantum effects. For the GS, many electrons are inside the quantum 
volume $a_B^d$ when $r_s$ is small and one gets a weakly coupled Fermi 
liquid. When $r_s$ is large, the volume per electron $a^d$ is large 
compared to $a_B^d$, and one gets an electron solid (Wigner crystal) 
with almost negligible quantum effects. Hereafter, though our theory 
could be easily extended, we restrict ourselves to polarized electrons 
(spinless fermions) at $d=2$. Assuming periodic boundary conditions 
(BCs) for a square of size $D$, one can ignore the constant term, 
the electronic density $n_s=N/D^2$ and $a=1/\sqrt{\pi n_s}$. 

 Let us now define a square lattice model of spacing 
$s$, size $L=D/s$, nearest neighbor hopping element 
$t=\hbar^2/(2m s^2)$ and interaction strength $U=e^2/s$. 
The lattice Hamiltonian $H_l$ reads:   
\begin{equation}
H_l=t \left(4N- \sum_{\left<{\bf j},{\bf j'}\right>}
c_{\bf j}^{\dagger} c_{\bf j'}\right) + \frac{U}{2}  
\sum_{{\bf j} \neq {\bf j'}} \frac{n_{\bf j} n_{\bf j'}}
{|d_{\bf jj'}|}.  
\label{H-lattice-site}
\end{equation}
The operators $c_{\bf{j}}^{\dagger}$ ($c_{\bf{j}}$) create 
(annihilate) a spinless fermion at the site $\bf{j}$ and 
$\left< {\bf j},{\bf j'}\right>$ means that the sum 
is restricted to nearest neighbors. $d_{\bf jj'}$ is the distance  
between the sites ${\bf j}$ and ${\bf j'}$ in unit of $s$.
When one takes periodic BCs, a convention has to be chosen for 
the distance $d_{\bf jj'}$. One possible definition of $d_{\bf jj'}$ is:
\begin{equation}
d_{\bf jj'}=\sqrt{\min(|j_x-j'_x|,L-|j_x-j'_x|)^2+
\min(|j_y-j'_y|,L-|j_y-j'_y|)^2}.
\label{distance1}
\end{equation}
Hereafter, we refer to the corresponding $1/|d_{\bf jj'}|$ repulsion as 
the periodic singular Coulomb (PSC) repulsion, since it has a cusp 
when the interparticle distance $d_{\bf jj'}$ has one of its coordinates 
equal to $L/2$.  This cusp being unphysical, we define also the 
periodic regularized Coulomb (PRC) repulsion defined from 
\begin{equation}
d_{\bf jj'}=\frac{L}{\pi} \sqrt{\sin^2\frac{|j_x-j'_x|\pi}{L}+
\sin^2\frac{|j_y-j'_y|\pi}{L}}
\label{distance2}
\end{equation}
which locally coincides with the PSC repulsion, but remains analytic 
for all values of $d_{\bf jj'}$ when $s \rightarrow 0$. The PRC repulsion 
is essentially equivalent to the Ewald repulsion obtained from the periodic 
repetition of the considered system. In the lattice units, $1 Ry=U^2/4t$ and 
the ratio $r_s$ becomes:      
\begin{equation}
r_s=\frac{UL}{2t\sqrt{\pi N}}.
\end{equation} 
 The question is to know whether $UL/(2t\sqrt{\pi N})$ 
remains a scaling parameter for $H_l$. The answer is positive if 
$r_s<r_s^*$. Below $r_s^*$, the lattice GS depends on the ratio $UL/t$ for 
a fixed value of $N$, and coincides with the continuum GS. 
Let us define three criteria giving $r_s^*$. 

{\it Criterion 1:} In the limit $t=0$, the $N$ electrons are localized 
on $N$ sites and form states $\left|J\right>=c_{\bf{j}_1}^{\dagger} \ldots 
c_{\bf{j}_N}^{\dagger} \left| 0 \right>$ 
of energy $E_{Coul}(J)$. As one turns on $t$, one can expect 
that the lattice becomes irrelevant as each electron ceases to be 
localized on a single site. In analogy with the problem of a single 
particle in a disordered lattice, one can use the criterion first 
proposed by Anderson \cite{anderson}: delocalization takes place when 
the hopping term $t$ between directly coupled sites becomes of the 
order of their energy spacing $\Delta E$. This criterion was extended 
to interacting systems in many different contexts: onset of quantum chaos 
in many body spectra \cite{wp,shepelyansky-suskhov,wpi} and in the quantum 
computer core \cite{benenti}, quasi-particle lifetime and delocalization 
in Fock space \cite{altshuler,jacquod}). In our case, the states 
become delocalized in the many body basis built from the states 
$\left|J\right>$ when the matrix element $\left<J'|H_1|J\right>$ of 
the one body perturbation $H_1 \propto t$ coupling a state 
$\left|J\right>$ to the ``first generation'' of states $\left|J'\right>$ 
directly coupled to it by $H_1$ exceeds their energy spacing 
$\Delta E_{Coul} = E_{Coul}(J')-E_{Coul}(J)$. 
This gives $t > \Delta E_{Coul}$. Applying this criterion to 
the GS, one obtains $r_s^*$ from the condition 
\begin{equation}
t \approx \Delta E_{Coul} (t=0),   
\label{Anderson}
\end{equation}     
where $\Delta E_{Coul}(t=0)$ is the first level spacing of the Coulomb 
system on a lattice. When $t$ exceeds $\Delta E_{Coul} (t=0)$, 
the GS is delocalized on the $J$-basis, and hence on the lattice, and 
the lattice GS becomes identical to the continuum GS. 

{\it Criterion 2}:  Since a continuum model is invariant under 
translations, the motion of the center of mass can 
be decoupled from the relative motions. Thus $H_c$ can be decomposed 
in two parts, one related to the center of mass motion which is 
independent of the interaction, while the second one contains only 
the relative motions and hence the interaction. This has a 
very important consequence for the persistent current $I$ driven 
by an enclosed Aharonov-Bohm flux $\phi$ in a continuum model: $I$ 
is independent of $r_s$ and keeps its non interacting value. For having 
the topology of a $2d$ torus enclosing $\phi$ along the $x$-direction, 
one takes the corresponding curled BC in this direction, keeping periodic 
BC in the $y$-direction. For a sufficient $r_s$, the electrons form a 
Wigner solid and the small relative motions cannot feel the BCs. In this 
limit, $I$ is just given by the center of mass motion, which is independent 
of $r_s$, and hence coincides with its non-interacting value. This point 
remains correct for small $r_s$, as it was proven for $1d$-rings 
\cite{muller-groeling,krive,burmeister} and observed for $d=2$ 
\cite{muller-groeling}. In contrast, since the previous decomposition 
into two parts does not necessary hold for $H_l$, $I \neq I(r_s=0)$ 
for a lattice when  $H_l$ and $H_c$ have different GSs. The decay of $I$ 
above $r_s^*$ (small $t/U$ at fixed $N$ and $L$) can be evaluated 
\cite{sw,ksp,np} by the leading contribution (of order $N$) $I_{l}^{(N)} (t/U)
\propto t(t/U)^{N-1}$ of a $t/U$ lattice expansion. The value of $r_s$ 
for which
\begin{equation}
I(U=0) \approx I_{l}^{(N)} (t/U)
\end{equation}
gives the criterion 2 for $r_s^*$ 
(see Fig. \ref{Fig2} lower left). Instead of $I(\phi)$, one can prefer 
to use the Kohn curvature $C_K=\partial^2 E_0/\partial \phi^2$ evaluated 
at $\phi=0$ or the GS energy change $\Delta_{T} E_0= 
E_0(\phi=0)-E_0(\phi=\pi)$ when the BC is twisted in the $x$-direction.

{\it Criterion 3}: When $t/U \rightarrow 0$, the leading correction 
to the Coulomb energy of $H_l$ is $4Nt$. Since the correction 
$E_{vib} (r_s)$ to the Coulomb energy coming from the zero point 
vibrational motion of the solid when the GSs of $H_l$ and $H_c$ are 
similar cannot exceed this lattice limit $4Nt$ (see Fig. 1), 
$r_s^*$ is also given by the condition 
\begin{equation}
E_{vib} (r_s) \approx 4Nt,
\label{Vibration} 
\end{equation}  
assuming that the values of the lattice parameters can yield a Wigner 
solid for $r_s < r_s^*$. 

 Let us calculate the quantities used in those criteria 
for the case of $N=3$ spinless fermions on a square lattice with 
either the PSC or PRC repulsions. A study of the case $N=2$ can be found 
in Ref.\cite{mp}. For $t=0$, as shown in the insets 
of Figs. 1 and \ref{Fig2}, the configurations minimizing 
the PRC and PSC Coulomb energies are different. Moving one particle by a 
single hop increases these energies by an amount 
\begin{equation}
\Delta E_{Coul}^{(PRC)} \approx  \frac{7\sqrt{2}\pi^{3}U}{12\sqrt{3}L^3} \ \ 
; \ \ \Delta E_{Coul}^{(PSC)} \approx  \frac{\sqrt{2}U}{L^2} 
\label{psc}
\end{equation} 
respectively when $L$ is sufficiently large. For $U=0$, the GS energy 
is given by $E_0(0)=12t-8t-4t\cos(2\pi/L)$ for periodic BCs and becomes 
$E_0(\pi)=12t-8t\cos(\pi/L) -4t\cos(3\pi/L)$ when one twists the 
BC in the $x$-direction. When $t/U$ is small, $\Delta_{T} E_0$ can be 
calculated at the leading order of a $t/U$-expansion \cite{np} 
for $N=3$. This gives when $L$ is large: 
\begin{equation}
\lim_{r_s \rightarrow 0} \Delta_{T} E_0 \approx \frac{14 \pi^2 t}{L^2} 
\ \ ;\ \ 
\lim_{r_s \rightarrow \infty} \Delta_{T} E_0 \approx \frac{9 \pi^2 
t^3}{L^2 \Delta E_{Coul}}
\label{current}
\end{equation}
where  $\Delta E_{Coul}$ is given by the Eqs. (\ref{psc}). 
Using these expressions, one obtains from the two first criteria:
\begin{equation}
r_s^*(L) = A L^{\alpha}
\label{threshold}
\end{equation}
where $\alpha =4$ for the PRC repulsion and $\alpha=3$ for the PSC repulsion, 
the constant $A$ slightly depending on the taken criterion.

For using the third criterion, one needs the zero point vibrational 
energy of the Wigner molecule that the three 
particles form for a sufficiently large $r_s<r_s^*$. Since the GSs of 
$H_l$ and $H_c$ are identical for $r_s<r_s^*(L)$, instead of $H_l$, 
one can use $H_c$ which is the sum of two decoupled terms. Denoting 
${\bf R}=(\sum_i^3 {\bf r}_i)/3$ the coordinate of the center of mass, 
the first term reads $H_{CM}=(\hbar^2/6m) \nabla^2_{\bf R}$ 
and corresponds to the rigid translation of the molecule while the other 
term contains the relative motions and the interaction. For a Wigner 
molecule, the second part can be simplified and expressed in terms of 
the normal coordinates suitable for describing the small vibrations 
around equilibrium. 

 The PRC repulsion is harmonic around equilibrium, and the three particles 
form a diagonal chain as indicated in the inset of Fig. 1 when $L/3$ is integer. One gets four decoupled harmonic oscillators, 
two corresponding to a longitudinal mode of frequency $\omega_l=\sqrt{20 B}$, 
the two others being a transverse mode of frequency $\omega_t=\sqrt{8B}$, 
where $B=(\sqrt{6} e^2 \pi)/(24 D^3 m)$. The zero point vibrational energy 
is then given for $N=3$ by: 
\begin{equation}
E_{vib} (r_s,N=3)= \hbar (\omega_l+\omega_t)=
2\pi \frac{\sqrt{5} +\sqrt{2}}{\sqrt{18}} \left(\frac{2}{\pi} \right)^{1/4} 
r_s^{-\beta}
\label{Vibration-3-PRC}
\end{equation} 
in rydbergs where $\beta=3/2$

 When one takes the PSC repulsion, the three relative distances 
at equilibrium are precisely ${\bf r}=(L/2,L/2)$, ${\bf r}=(0,L/2)$ and 
${\bf r}=(L/2,0)$ respectively when $L$ is even. The potentials 
$v(\delta{\bf r})$ felt by the electrons around their equilibrium positions 
are singular and can be expanded as $v(\delta{\bf r}) \approx C_1 
|\delta r_x|+ C_2|\delta r_y|$, where $C_1$ and $C_2$ depend on the 
equilibrium positions and are $\propto e^2/D^2=U/L^2$. For a single 
particle in a $1d$-potential $v(x)=C|x|$, the GS energy $\epsilon$ can be 
approximated by $t/B^2+CB$ where $B$ is the GS extension and  
is given by $\partial \epsilon/\partial B=0$. This yields 
$B\propto (C/t)^{1/3}$ and $\epsilon \propto (U^2t/L^4)^{1/3}$. 
Since the $2d$-potential 
$v(\delta{\bf r})$ is separable, one eventually finds: 
\begin{equation}
E_{vib} (r_s,N=3) \propto r_s^{-\beta}
\label{Vibration-3-PSC} 
\end{equation}
in rydbergs where $\beta=4/3$. As one can see, the PSC repulsion 
gives a higher exponent $\beta$ when $N=3$, which is inconsistent with 
the usual expansion \cite{wigner} in powers of $r_s^{-1/2}$ 
first proposed by Wigner. 

From Eq. (\ref{Vibration-3-PRC}) and Eq. (\ref{Vibration-3-PSC}), 
since $4Nt$ in rydbergs is given by $(4L^2)/(\pi r_s^2)$, one 
obtains from the third criterion similar results for $r_s^*$ as 
in Eq. (\ref{threshold}), with $\alpha=4$ 
for the PRC repulsion and $\alpha=3$ for PSC repulsion. 

\begin{figure} 
\vspace{0.5cm}
\begin{center}
%\centerline{
\epsfxsize=10cm 
\epsfbox{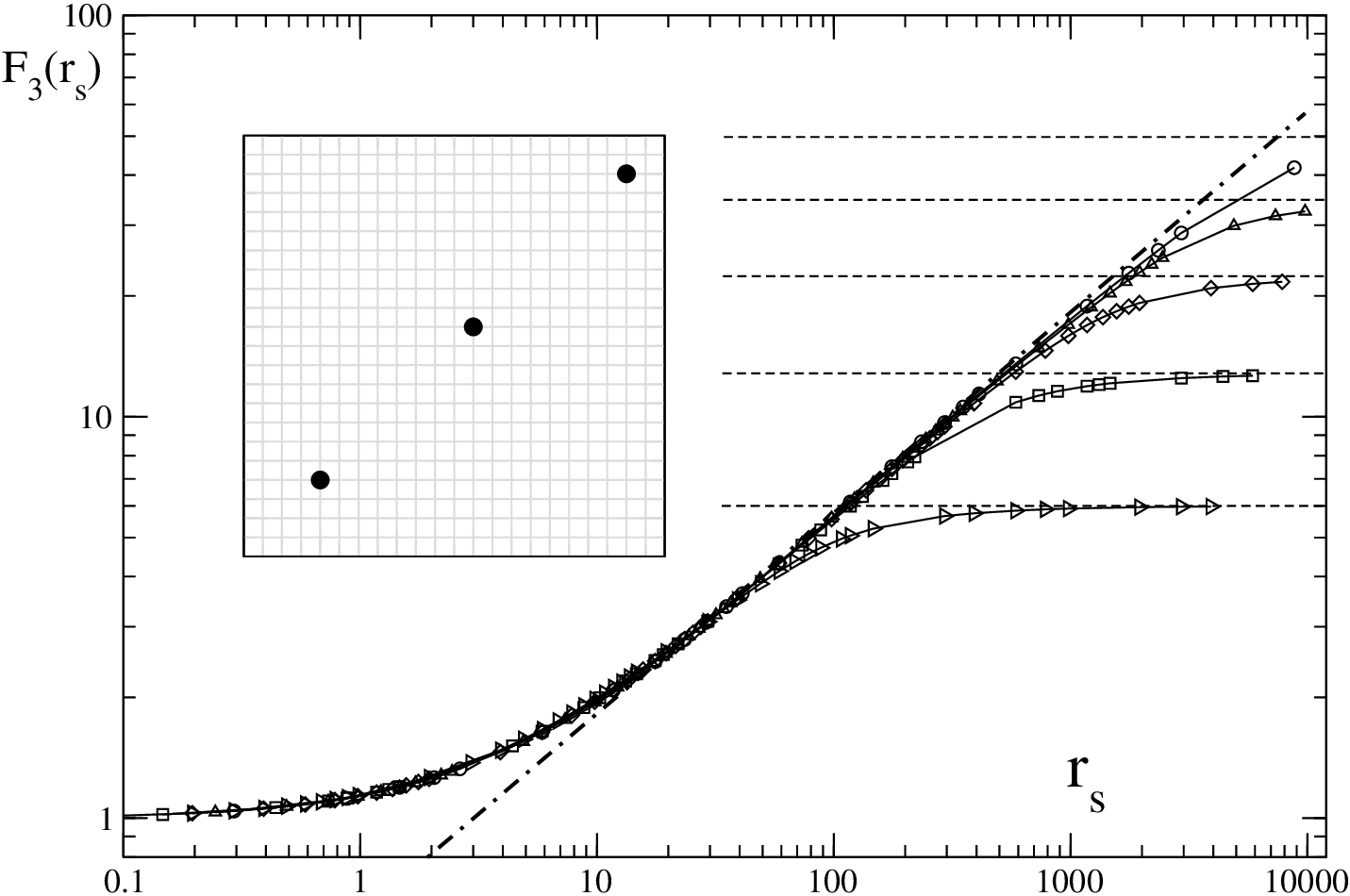}
%\epsffile{figure1.eps}
%}
\vspace{0.5cm}
\caption
{
Energy ratio $F_{N=3}(L,U,t)$ as a function of 
$r_s=(UL)/(2t \sqrt{\pi N})$ given by the PRC repulsion 
for $L=6$ ($\triangleright$), 9 ($\Box$), 12 ($\diamond$), 15 
($\triangle$), 18 ($\circ$). The dotted-dashed line gives the 
behavior $0.5764 \sqrt {r_s}$ (harmonic vibrations of the 
Wigner molecule) and intersects the limiting dashed lines 
$12t/(4t-4t\cos(2\pi/L))$ at the $r_s^*(L)$ corresponding to 
criterion 3. Inset: A GS configuration when $t=0$ and $L=24$.
}
\end{center}
\label{Fig1}
\end{figure}

 We now present numerical results obtained using the Lanczos 
algorithm, after having written $H_l$ in terms of the 
creation (annihilation) operators of a single particle in a state 
of momentum ${\bf k}$ and having calculated the GSs in the sub-space 
of total momentum ${\bf K}=0$ \cite{np}. From the GS energy $E_0(L,U,t)$ of 
${\bf K}=0$, and for a given value of $N$, we define the dimensionless 
ratio $F_N(L,U,t)$ by: 
\begin{equation} 
F_N(L,U,t)=\frac{E_0(L,U,t)-E_0(L,U,t=0)}{E_0(L,U=0,t)}.
\label{ratio}
\end{equation} 

The results for the PRC repulsion are shown in Figure 1. 
For $t=0$, the values of $L=6,9,12,15,18$ are 
commensurate with the period of the diagonal Wigner molecule 
shown in the inset, reducing the lattice effects. 
When $F_{N=3}(L,U,t)$ is plotted as a function of $r_s= 
(UL)/(2t \sqrt{\pi N})$, the different functions $F_{N=3}(L,U,t)$ 
scale without an observable lattice effect up to the $r_s^*(L)$ 
exactly given by criterion 3. Using Eq. (\ref{Vibration-3-PSC}) and 
$E_0(L,U=0,t)=12t-8t-4t\cos(2\pi/L)$ one can see that the numerical 
results coincide with the analytical result $F_{N=3} = 
0.5764 \sqrt{r_s}$ valid for a continuum Wigner molecule. The 
function  $F_{N=3}(L,U,t)$ saturates to $4Nt/E_0(L,U=0,t)$ above 
$r_s^*(L)$, as indicated by the dashed lines.

\begin{figure}
\begin{center}
\epsfxsize=6.5cm 
%\epsffile{figure2.eps}
\epsfbox{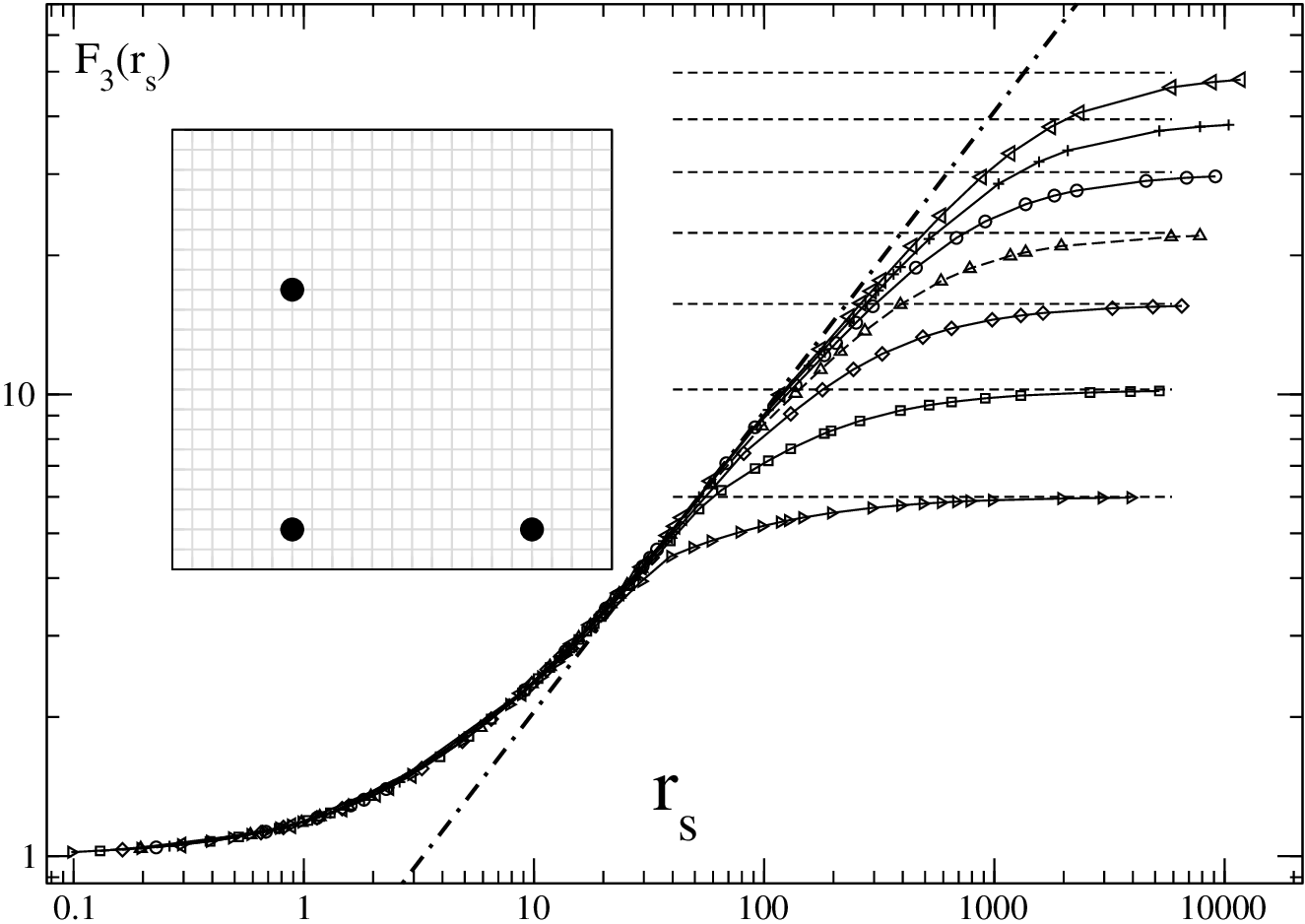}
\qquad
\epsfxsize=6.5cm 
\epsfbox{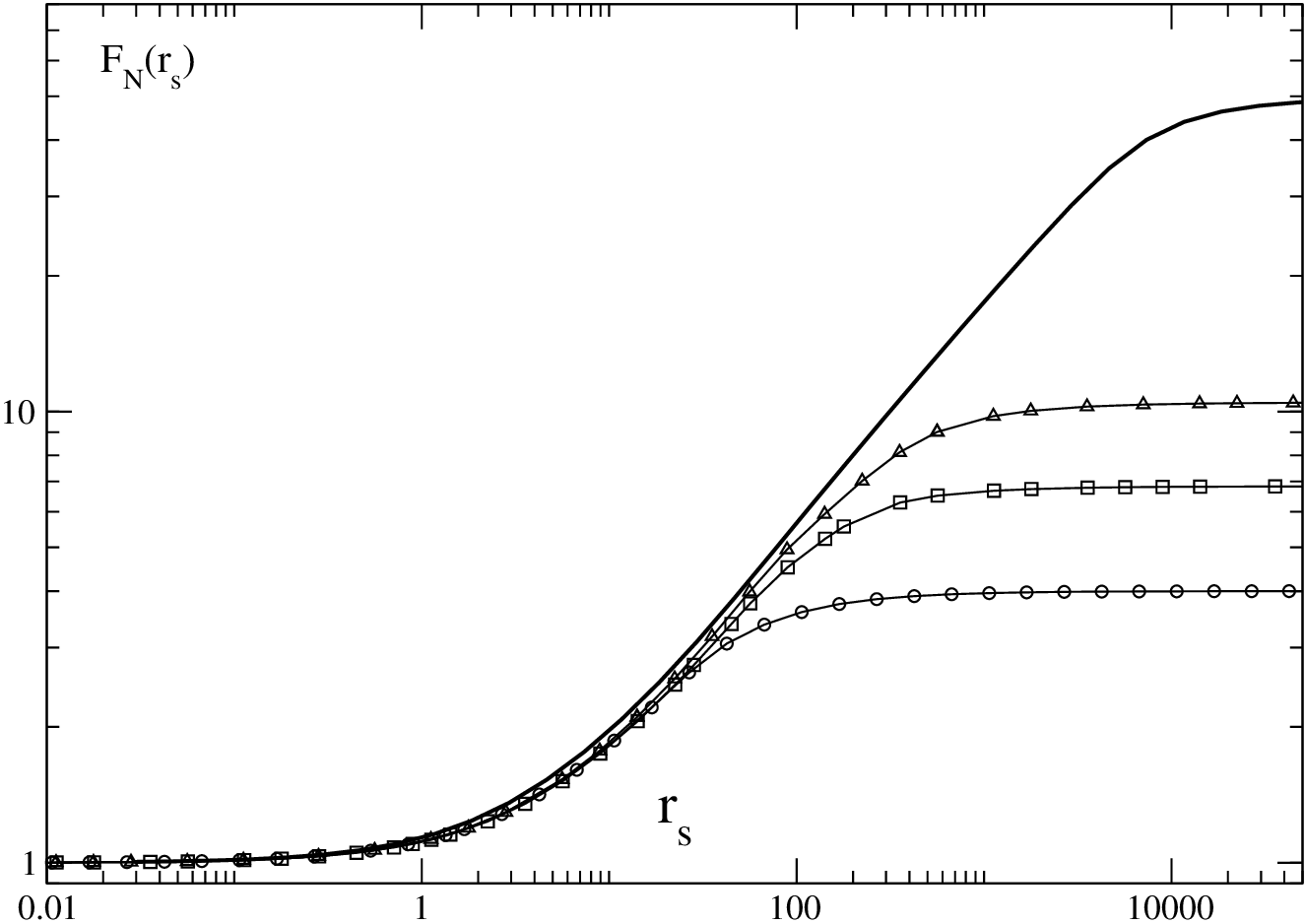}
%\epsffile{figure3.eps}
\end{center}
\end{figure}
\begin{figure}
\begin{center}
\epsfxsize=6.5cm 
\epsfbox{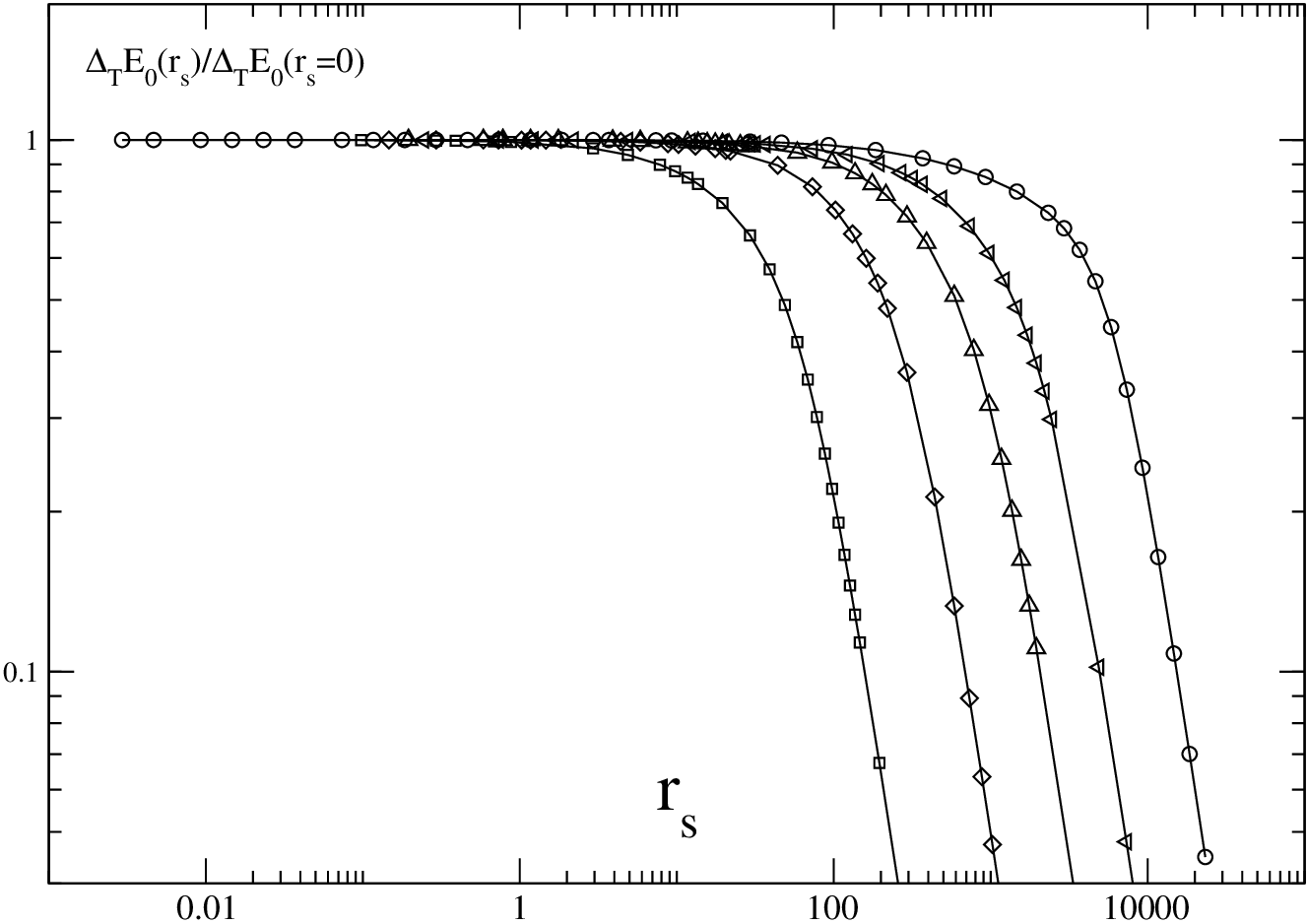}
%\epsffile{figure4.eps}
\qquad
\epsfxsize=6.5cm 
\epsfbox{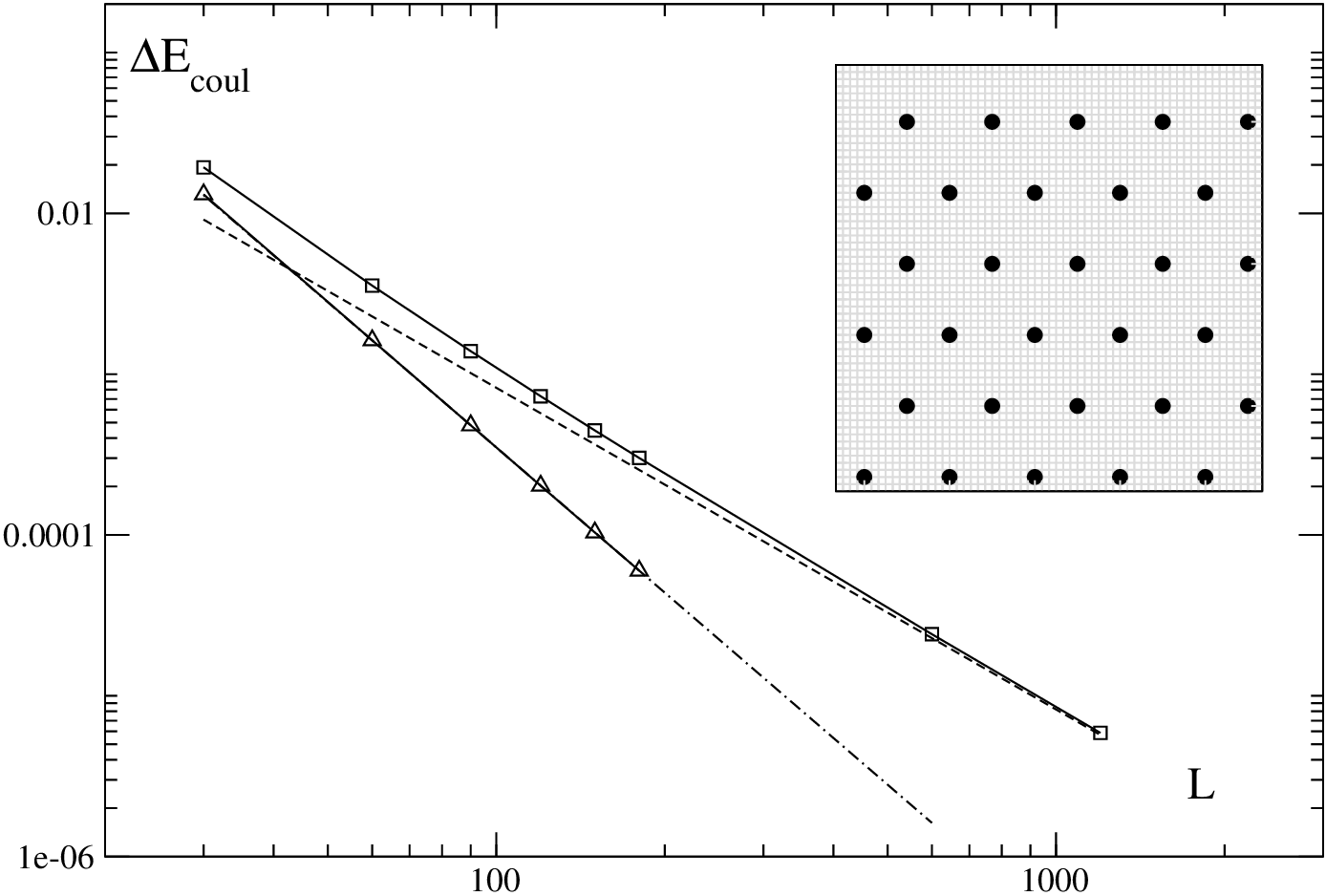}
%\epsffile{figure5.eps}
\end{center}
%\vspace{1cm}
\caption
{Upper left: Same as in Fig. 1 using the PSC repulsion, 
where the dotted-dashed line now gives the $r_s^{2/3}$ behavior 
due to the vibrations of the Wigner molecule.
Inset: a GS configuration when $t=0$ and $L=24$. 
Upper right: Energy ratios $F_{N} (L,U,t)$ using the PRC repulsion 
for $N=3$ ($L=18$ solid line) and $N=4$ ($L=6$ $\circ$, 8 $\Box$, 10 
$\triangle$) as a function of $r_s=UL/(2t\sqrt{\pi N})$.  
Lower left: Dimensionless change $\Delta_T E_0 (r_s)/ 
\Delta_T E_0 (r_s=0)$ of the GS energy when the longitudinal BC is 
twisted for $L=6$ ($\Box$), 9 ($\diamond$) ,12 ($\triangle$), 15 
($\triangleleft$), 18 ($\circ$) as a function of $r_s$ (PRC repulsion). 
Lower right: First energy spacing $\Delta E_{Coul}$ 
yielded by the hop of one particle from the $t=0$ GS as a 
function of $L$. $L^{-3}$ behavior yielded by the PRC repulsion 
($\triangle$) for $N=30$. Crossover towards the $L^{-2}$ decay 
(dashed line) yielded by the PSC repulsion ($\Box$). Inset: 
One GS configuration yielded by the two repulsions ($t=0$, 
$N=30$ and $L=30$). 
}
\label{Fig2}
\end{figure}

The corresponding results for the PSC repulsion are shown in Fig.  
\ref{Fig2} (upper left) for even values of $L$, where the GS is a 
triangular molecule shown in the inset when $t/U \rightarrow 0$.  
Again the curves scale up to the onset $r_s^*(L)$ given by criterion 
3. But, $F_{N=3} \propto r_s^{2/3}$ for intermediate $r_s$, as 
implied by Eq. (\ref{Vibration-3-PRC}), and not $\propto r_s^{1/2}$. 
 
 In Fig. \ref{Fig2} (upper right), a small change of the scaling curve 
$F_{N}$ can be seen when a fourth electron is added, accompanied 
by the expected breakdown of the  scaling behavior above $r_s^*$. 
When $N \rightarrow \infty$, $F_{N}$ should converge towards a 
thermodynamic limit depending only on $r_s$. Unfortunately, a study 
of this convergence is out of reach of a numerical approach using 
exact diagonalization. 

 In Fig. \ref{Fig2} (lower left), we illustrate the criterion 2,  
showing for $N=3$ the change $\Delta_{T} E_0$ of the GS energy 
when a BC is twisted. Below $r_s^*$, $\Delta_{T} E_0$ does not 
depend on the interaction, while above $r_s^*$, one gets 
the lattice limit given by Eq. (\ref{current}). We have checked that 
the same conclusions can be drawn from a study of the Kohn curvature 
$C_K(r_s)$. 

 The PSC and PRC repulsions give rise to different $r_s^*(L)$, 
$F_{N=3} \propto r_s^{2/3}$ for the PSC repulsion, differing from 
the conventional expansion in powers of $r_s^{1/2}$. 
Does this difference remain for larger values of $N$? Indeed the 
contribution of pairs $ij$ having the coordinates of their spacings 
$d_{ij}$ close to $D/2$, and responsible for the $r_s^{2/3}$ behavior 
when $d_{ij}$ is defined by Eq. (\ref{distance1}), becomes a surface 
effect $\propto N$ compared to the bulk contribution $\propto N^2$ 
of the remaining pairs, yielding $\Delta E_{Coul}^{PSC} \approx AN/L^2 
+ BN^2/L^3$, where $A$ and $B$ are constant. For a fixed $L$ and 
increasing $N$, $\Delta E_{Coul}^{PSC} \rightarrow B N^2/L^3$ and following 
criterion 1, the conventional $r_s^{1/2}$ expansion for $F_{N}$ should 
valid for the PSC repulsion too. However, for $N$ fixed and 
increasing $L$, the surface contribution dominates and the difference 
between the two repulsions remains in this limit. 
To illustrate this point, we have studied a Coulomb system of $N=30$ 
electrons. The electron configuration minimizing the Coulomb energy 
is shown in the inset of Fig. \ref{Fig2} (lower right) for $L=30$, 
i.e. one of the values of $L$ for which the hexagonal electron lattice 
is commensurate with the underlying lattice. This configuration is 
given by the PRC and PSC repulsions. As shown in Fig. \ref{Fig2} 
(lower right) $\Delta E_{Coul}^{PSC}$ is in the crossover regime 
between the bulk regular $\propto L^{-3}$ and the singular surface 
$\propto L^{-2}$ behaviors, while one has the expected $L^{-3}$ for 
$\Delta E_{Coul}^{PRC}$.

%%%%%%%%%%%%%%%%%%%%%%%%% Conclusion %%%%%%%%%%%%%%%%%%%%%%%%%%%%%%%%%%%
 
 We summarize the main results which we have checked using small 
lattice models and varying either $U/t$ or $L$ for $N=2,3,4$. For 
$r_s < r_s^*(L)$, the GSs of $H_l$ and $H_c$ are identical, 
$r_s$ is the scaling parameter and $I(r_s)=I(r_s=0)$. Above $r_s^*(L)$, 
$r_s$ ceases to be the scaling parameter and $I$ decays. The following 
relations have been obtained independently of the definition of the 
Coulomb repulsion: $r_s^* \propto L^{\alpha}$, $E_{vib} 
\propto r_s^{-\beta}$  and $\Delta E_{Coul} \propto UL^{-\gamma}$, with 
the relations $\alpha= \gamma+1$ and $\alpha =2/(2-\beta)$ 
between the exponents. 

Our motivation was to investigate whether the study of a lattice model 
is relevant for a continuum model, fixing $N$ and varying the 
lattice parameters. Another issue, of more direct physical relevance, 
is to study the role played by the existing lattice upon a $2d$ gas of 
correlated conduction electrons. We postpone the discussion of this 
question, where the lattice parameters are given, while $N$ can be varied, 
to a following work.

%%%%%%%%%%%%%%%%%%%%%%%%%%%%%%%%%%%%%%%%%%%%%%%%%%%%%%%%%%%%%%%%%%%

%
%++++++++++++++++++acknowledgment+++++++++ 
%

 We thank X. Waintal for his help in the study of a Coulomb system of 
$N=30$ electrons. One of us (Z\'AN) acknowledges the support of the European 
Community's Human Potential Programme under contract HPRN-CT-2000-00144 
and the Hungarian Science Foundation OTKA TO34832.

\end{document}